\documentclass[twocolumn,aps,prl,showpacs,,showkeys,groupedaddress,amsmath,amssymb]{revtex4}
\usepackage{amsmath}
\usepackage{amssymb}
\usepackage{txfonts}
\usepackage{graphicx}
\usepackage{dcolumn}
\begin{document}
\title{Quantum Interfaces Using Nanoscale Surface Plasmons}
\author{Fang-Yu Hong}
\author{Shi-Jie Xiong }

\affiliation{National Laboratory of Solid State Microstructures and
Department of Physics, Nanjing University, Nanjing 210093, China}
\date{\today}
\begin{abstract}
The strong coupling between individual optical emitters and
propagating surface plasmons confined to a conducting nanotip make
this system act as an ideal interface for quantum networks, through
which a stationary qubit and a flying photon (surface plasmon) qubit
can be interconverted via a Raman process. This quantum interface
paves the way  for many essential functions of a quantum network,
including sending, receiving, transferring, swapping, and entangling
qubits at distributed quantum nodes as well as a deterministic
source and an efficient detector of a single-photon. Numerical
simulation shows that this scheme is robust against experimental
imperfections and has high fidelity. Furthermore, being smaller this
interface would significantly facilitate the scalability of quantum
computers.

\end{abstract}



\maketitle
\section {I. INTRODUCTION}
Quantum networks comprised of local nodes and quantum channels are
of fundamental  importance  for quantum communication and essential
for scalable and distributed quantum computation \cite{jcae, ddiv}.
A quantum interface mapping between local and flying qubits is the
key component of the quantum network. Schemes for this purpose
utilizing strong coupling  \cite{rtgr,mbru}  between a high-Q
optical cavity and the atoms have been suggested
\cite{cirac,adia,wang}. However, such schemes put challenging
constraints on optical cavities which is difficult to be
miniaturized. A novel scheme based on surface plasmons (SPs) to
reach the strong-coupling regime on a chip has been intensively
explored \cite{lcas, assr, abla, awal, dcas,dcass,mist,crcn}. A
substantial increase in the coupling strength
$g\propto1/\sqrt{V_{eff}}$ can be achieved through using SPs where
the effective mode volume $V_{eff}$ for the photons can be greatly
reduced \cite{dcass}.  An effective Purcell factor
$P\equiv\Gamma_{p}/\Gamma' \approx 2.5\times 10^3$ in realistic
systems is possible \cite{dcasp}, where $\Gamma_{p}$ is the
spontaneous emission rate into the surface plasmons and $\Gamma'$
describes the loss rate into other channels. Furthermore, unlike the
strong coupling based on  cavity quantum electrodynamics (QED), this
strong coupling is broadband \cite{dcass}.

On account of these considerations, we  proposes a general control
scheme of emitter-photon quantum interface based on the strong
interaction between surface plasmons in a nanotip and an optical
emitter. The process of the state transfer between two nodes can be
separated into two steps: the sending operation at one node mapping
a stationary qubit into a flying qubit and the receiving operation
at another node mapping the flying qubit into  a stationary one.
With the advance in the pulse shaping technique \cite{dsss}, two
aspects of the process are controllable: the production of an
arbitrarily shaped pulse under the condition that it is sufficiently
smooth and the operation of the Raman process as a partial cycle, in
which the initial state $|s\rangle|vac\rangle$ is mapped into an
entangled state
$\cos\theta|s\rangle|vac\rangle+\sin\theta|g\rangle|E(t)\rangle$ for
any $\theta\in[0,\pi/2]$, where $|g\rangle$ and $|s\rangle$ are the
stationary qubit states and the flying qubit is denoted by the
vacuum state $|vac\rangle$ and a single plasmon (photon) state with
wave packet $|E(t)\rangle$.

 A number of essential functions of a quantum network can be fulfilled by this quantum
interface:(1) It can send a flying quantum state and can also
function as a deterministic source of a single photon with arbitrary
pulse shape and controllable average photon number. (ii) It can
receive a flying quantum state \cite{inag}, being an efficient
single-photon detector if the incoming photon pulse shape is known.
(iii) A state can be transferred from one node to another. (iv) An
entanglement between either two remote stationary qubits or  a
stationary qubit and a flying qubit can be generated in a partial
Raman cycle. Numerical simulations of this scheme demonstrate
robustness against parameters errors and high fidelity.
 With stronger emitter photon coupling strength,
 faster manipulation times can be expected.
  Furthermore, as the setup is much easier to be made smaller,
this scheme would open the possibility to higher scalability of
quantum computers.
\section {II. EXACT SOLUTION OF THE QUANTUM INTERFACE DYNAMICS}
\begin{figure}
\includegraphics[scale=0.5]{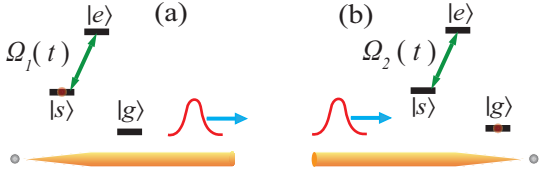}
\caption{\label{fig:1} (Color online) Schematic picture of the
interface comprised of an optical emitter and a nanotip for the
quantum network. (a) The interface for sending process. A
three-level optical emitter starts in state $|s\rangle$, is coupled
to excited state $|e\rangle$ via a time-dependent external control
field $\Omega_1(t)$. We assume that the excited state $|e\rangle$ is
coupled to state $|g\rangle$ via the plasmon modes, causing
$|e\rangle$ to decay into $|g\rangle$ with high probability, while
simultaneously generating a single-photon in the plasmon modes. The
control pulse $\Omega_1(t)$ is determined by the  shape of the
generated photon wave packet which can be arbitrarily specified. (b)
The interface for receiving process. The emitter is initially in the
ground state $|g\rangle$. Under the action of $\Omega_2(t)$ which is
determined by the wave packet of the incoming photon, the emitter
can absorb a photon  while inducing a state flip from $|g\rangle$ to
  $|s\rangle$.  }
\end{figure}

The prototype quantum interface consists of a nanotip and a
three-level emitter  \cite{mist,crcn,dcasp,dcas} described by the
operator $\sigma_{ij}=|i\rangle\langle j|$, $i,j=e,g,s$
(Fig.\ref{fig:1}). Here, the qubit is represented by a ground state
$|g\rangle$  and a metastable state $|s\rangle$. State $|s\rangle$
is decoupled from the surface plasmons owing to, for example, a
different orientation of its associated dipole moment\cite{dcass},
but is resonantly coupled to excited $|e\rangle$ via some classical,
optical control field $\Omega(t)$ with central frequency $\omega_L$.
States $|g\rangle$ and $|e\rangle$ is coupled with strength $g$ via
the surface plasmon modes with wave vector $k$ which is  described
by annihilation operation $a_k$. States $|g\rangle$, $|s\rangle$,
and $|e\rangle$ have the energy $\omega_g=0$, $\omega_s$, and
$\omega_e$, respectively. The laser light satisfies the resonance
condition: $\omega_L+\omega_s=\omega_e$. Since the coupling g is
broad-band, it can be assumed to be frequency-independent
\cite{dcass,dcasp}. A linear dispersion relation $\omega _k=c|k|$
holds provided $\hbar\omega_k<2$ eV \cite{crcn,gsjk}, with $c$
denoting the group velocity of the SPs. Then similar to the
Hamiltonian in \cite{dcass} describing the interaction of  an
emitter and a nanowire,  the Hamiltonian for our model can be
written in the form
\begin{eqnarray}\label{eq1}
H&=&(\omega_{e}-i\frac{\Gamma'}{2})\sigma_{ee}+\omega_s\sigma_{ss}-(\Omega(t)e^{-it\omega_L}\sigma_{es}+H.c.)\notag\\
&+&\int_{-\infty}^{\infty} d k c|k|a_k^\dagger
a_k-(g\int_{-\infty}^{\infty} d k \sigma_{eg} a_k+H.c.),
\end{eqnarray}
where the emitter is assumed to be in the origin of an axis  and the
non-Hermitian term in $H$ describes the decay of state $|e\rangle$
at a rate $\Gamma'$ into all other possible channels \cite{dcas}.
This effective hamiltonian is in effect under the condition that
$k_B T\ll\hbar\omega_{e}$, e.g., if $\hbar\omega_{e}=1$ meV, $T<1$
K, where $k_B$ is the Boltzmann constant \cite{dcass}.

Note that the system described by this Hamiltonian, has two
invariant Hilbert subspaces, with the bases $\{|g,vac\rangle\}$ and
$\{|s,vac\rangle, |e,vac\rangle, |g,k\rangle\}$, respectively (
where in $|m,n\rangle, m=g,s,e$ denotes the state of three-level
system and $|k\rangle$ denotes the one-photon Fock state of the
surface plasmon mode of wave vector $k$). So the evolution of the
system containing one excitation can be generally described by the
state
$|\Psi(t)\rangle=\alpha_g|g,vac\rangle+\alpha_s|\psi(t)\rangle$,
where
\begin{equation}\label{eq8}
|\psi(t)\rangle=\int_{-\infty}^{\infty} d k
\beta_k(t)a_k^\dagger|g,vac\rangle+\beta_e(t)|e,vac\rangle+\beta_s(t)|s,vac\rangle.
\end{equation}
Under the Hamiltonian given in equation \eqref{eq1}, the equations
of motion  for the resonant Raman process (in a rotation frame) can
be derived as
\begin{equation}\label{eq15}
\dot{\beta}_k(t)=-i\delta_k \beta_k(t)+ig\beta_e(t),
\end{equation}
\begin{equation}\label{eq16}
\dot{\beta}_e(t)=-\frac{\Gamma'}{2}\beta_e(t)+i\Omega(t)\beta_s(t)+ig\int_{-\infty}^{\infty}
dk \beta_k(t).
\end{equation}

where $\delta_k=c|k|-\omega_e$. Integrating  equation \eqref{eq15}
yields
\begin{subequations}\label{eq17}
\begin{eqnarray}
\beta_k(t)&=&\beta_k(-\infty)e^{-i\delta_kt}+ig\int_{-\infty}^tdt'\beta_e(t')e^{-i\delta_k(t-t')},\label{eq2}\\
&=&\beta_k(\infty)e^{-i\delta_kt}-ig\int^{\infty}_tdt'\beta_e(t')e^{-i\delta_k(t-t')}\label{eq3}.
\end{eqnarray}
\end{subequations}

Equations \eqref{eq2} can be used for the receiving process, while
equation \eqref{eq3} for the sending process. Substituting equationspra57pra57
\eqref{eq17} into equation \eqref{eq16}, within the Wigner-Weisskopf
approximation \cite{mpsm,dcass}, we have the equations of motion for
the atomic state:

\begin{subequations}\label{eq7}
\begin{eqnarray}
\dot{\beta}_e(t)&=&i\Omega(t)\beta_s(t)-\frac{\Gamma_{p}+\Gamma'}{2}\beta_e(t)+i\sqrt{2\pi}gE_{in}(t),\label{eq4}\\
&=&i\Omega(t)\beta_s(t)-\frac{-\Gamma_{p}+\Gamma'}{2}\beta_e(t)+i\sqrt{2\pi}gE_{out}(t),\label{eq5}\\
\dot{\beta}_s(t)&=&i\Omega^\ast(t)\beta_e(t)\label{eq6},
\end{eqnarray}
\end{subequations}
where $\Gamma_{p}=2\pi g^2/c$ is the spontaneous emission rate into
the SP modes,
\begin{eqnarray}
E_{in}(t)&=&1/\sqrt{2\pi}\int_{-\infty}^{\infty} dk
\beta_k(-\infty)e^{-i\delta_k t}\notag
\\ &=&1/\sqrt{2\pi}\int_0^{\infty}
dk \beta_k(-\infty)e^{-i\delta_k t},
\end{eqnarray}
 and
$E_{out}(t)=1/\sqrt{2\pi}\int_{-\infty}^0 dk
\beta_k(\infty)e^{-i\delta_k t}$ are the incoming and outgoing
single-photon wave functions (in a rotating frame), respectively. We
have assumed that $\beta_k(-\infty)=0$ if $k<0$ for the incoming
field and $\beta_k(\infty)=0$ if $k>0$ for the outgoing field.

Below we will show that, from equations \eqref{eq7}, the amplitudes
$\beta_e(t)$ and $\beta_s(t)$ including the control pulse
$\Omega(t)$ can be expressed in terms of $E_{in}(t)$ and
$E_{out}(t)$. Thus the desired operation, with $E_{in}(t)$ and
$E_{out}(t)$ arbitrarily specified, can be generated on demand as
long as the normalization of the wave function of equation
\eqref{eq8} is satisfied.  From equations \eqref{eq4},\eqref{eq5},
we have
\begin{equation}
\beta_e(t)=\frac{ic}{\sqrt{2\pi}g}(E_{in}-E_{out}).
\end{equation}
From equations \eqref{eq7}, we can solve for the amplitude of
$\beta_s(t)$:
\begin{eqnarray}\label{eq9}
\frac{d}{dt}|\beta_s(t)|^2&=&c(|E_{in}(t)|^2-|E_{out}(t)|^2)-\frac{c}{P}|E_{out}(t)-E_{in}(t)|^2\notag\\
&-&\frac{c}{\Gamma_{p}}\frac{d}{dt}|E_{out}(t)-E_{in}(t)|^2
\end{eqnarray}
and the phase:
\begin{eqnarray}\label{eq10}
\frac{d\theta}{dt}&=&\frac{i}{|\beta_s(t)|^{2}}\left[\beta_e(t)\left(\frac{d}{dt}\beta^\ast_e(t)
+\frac{\Gamma_{p}+\Gamma'}{2}\beta^\ast_e(t)\right.\right.\notag\\
&+&\left.\left.i\sqrt{2\pi}gE_{in}^\ast(t)\right)+\frac{1}{2}\frac{d}{dt}|\beta_s(t)|^2\right].
\end{eqnarray}
Then, from equation \eqref{eq6}, we can express $\Omega(t)$ in terms
of the amplitudes that have been solved above:
\begin{equation}\label{eq11}
\Omega(t)=i\left(\frac{d}{dt}\beta^\ast_s(t)\right)/\beta^\ast_e(t).
\end{equation}

For the sending node of the quantum network, the initial conditions
are $E_{in}(t)=0$, $\beta_e(-\infty)=0$, and $\beta_s(-\infty)=1$.
The outgoing single-photon wave packet can contain average
$\sin^2\theta$ photon: $\int_{-\infty}^\infty
d\tau|E_{out}(\tau)|^2=\sin^2\theta\int_{-\infty}^\infty d\tau
|\tilde{E}_{out}(\tau)|^2=\sin^2\theta/c$, where
$\tilde{E}_{out}(t)$ is the normalized wavepacket of the emitted
photon. At the remote future time $t\rightarrow+\infty$, the photon
emission process is completed, we have $\beta_e(t)=0$,
 and $\beta_s(t)=(1-(1-1/P)\sin^2\theta)^{1/2}e^{i\phi}\approx \cos\theta
e^{i\phi}$, with the controllable phase given by equation
\eqref{eq10}. The most general form of the photon generation process
can be described by
\begin{eqnarray}\label{eq12}
\alpha_g|g,vac\rangle&+&\alpha_s|s,vac\rangle\xrightarrow{\Omega(t)}\alpha_g|g,vac\rangle\notag
\\&+&\alpha_s[e^{i\phi}\cos\theta|s,vac\rangle
+\sin\theta|g,\tilde{E}_{out}(t)\rangle]
\end{eqnarray}
If $\theta=\pi/2$ and equation \eqref{eq12} is reduced to the
equation
\begin{equation} \label{eq13}
(\alpha_g|g\rangle+\alpha_s|s\rangle)|vac\rangle\rightarrow|g\rangle(\alpha_g|vac\rangle+\alpha_s|\tilde{E}_{out}(t))\rangle.
\end{equation}
 mapping the stationary qubit onto the flying qubit.
Further, if initially the emitter is entirely in state $|s\rangle$,
then this mapping operation can work as the deterministic generation
of a single-photon wave packet with any desired pulse shape
$\tilde{E}_{out}(t)$. If $\theta<\pi/2$, this sending node can also
function as generation of entanglement between the emitter and the
flying qubit:
\begin{equation}
|s,
vac\rangle\xrightarrow{\Omega(t)}e^{i\phi}\cos\theta|s,vac\rangle+\sin\theta|g,\tilde
{E}_{out}(t)\rangle.
\end{equation}

The receiving process is basically the time reversal of the
full-cycle
 sending process. With the emitter initially in state
$|g\rangle$ and the incoming flying qubit
$\alpha_g|vac\rangle+\alpha_s|\tilde {E}_{in}(t)\rangle$, the
mapping transformation is expressed by
\begin{equation}\label{eq14}
|g\rangle(\alpha_g|vac\rangle+\alpha_s|\tilde{E}_{in}(t)\rangle)\rightarrow(\alpha_g|g\rangle+\alpha_s|s\rangle)|vac\rangle.
\end{equation}
 As in the sending process,
the incoming photon pulse $\tilde{E}_{in}(t)$ photon can be
arbitrarily specified, provided that it is smooth enough and without
the outgoing photon. As the stationary qubit can be read out
non-destructively \cite{lryw,mtag}, the receiving node can also
function as a photon detector when the photon pulse shape is known.

By combining the sending and receiving process, the transfer of
qubit from one node to another can be easily accomplished. When two
state transfer operations with opposite directions are performed at
the same time, the two qubits are swapped. If $\theta<\pi/2$, the
joint operation of the sending and receiving process can produce an
entangled state of the two nodes by the transformation:
\begin{equation}
|s,g\rangle|vac\rangle\xrightarrow[\Omega_2(t)]{\Omega_1(t)}\left(e^{i\phi}\cos\theta|s,g\rangle+\sin\theta|g,s\rangle\right)|vac\rangle.
\end{equation}

\begin{figure}
\includegraphics[scale=0.35]{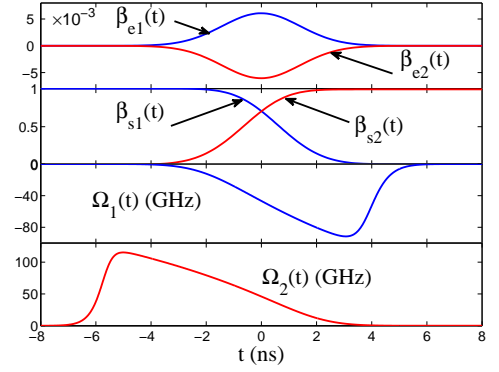}
\caption{\label{fig:2} (Color online) Transfer of a  qubit from the
sending node to the receiving node. (a) Amplitudes of the state
$\beta_{e1}$ and $\beta_{e2}$. (b) Amplitude of the states
$\beta_{s1}$ and $\beta_{s2}$. (c) The control field $\Omega_1(t)$
for the sending process. (d) The control field $\Omega_2(t)$ for the
receiving process.}
\end{figure}

Before surface plasmons decay they can travel about 140 plasmon
wavelengths \cite{dcas} which corresponding to about 0.2 m for the
energy split $\omega_{eg}=1$ mev and dielectric permittivity
$\epsilon=50$ \cite{dcas}. Thus, the loss of the SP during the
travel can be negligible if the two node is about 1 $\mu$m apart
from each other. In our model, the difference in energy levels
between  the two  emitters in the two nodes can be allowed to exist,
thus in the realistic systems, the two emitter can be independently
addressed and  the effect that a close spacing between two atom
affects the spontaneous emission process \cite{rhle} can be
suppressed. Short distance quantum communications are essential for
a quantum computer. For long distance communications, the SPs  can
be in- and out-coupled to conventional waveguides \cite{dcass}.

\section {III. NUMERICAL SIMULATIONS}
\begin{figure}
\includegraphics[scale=0.35]{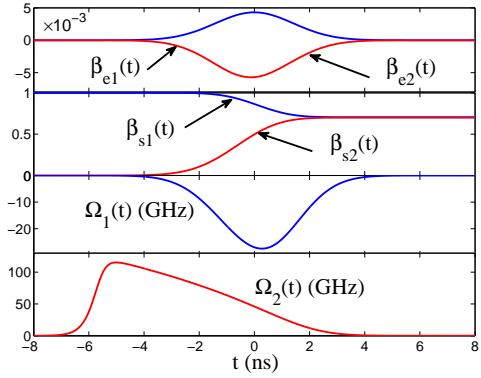}
\caption{\label{fig:3} (Color online) Generation of entanglement of
two qubits in neighboring nodes with the target state:
$(|s,g\rangle+|g,s\rangle)/\sqrt{2}$. (a) Amplitude of the states
$\beta_{e1}$ and $\beta_{e2}$. (b) Amplitude of the states
$\beta_{s1}$ and $\beta_{s2}$. (c) The driving field $\Omega_1(t)$
for the sending process. (d) The driving field $\Omega_2(t)$ for the
receiving process.}
\end{figure}

In the following numerical simulations, for simplicity we assume
that $E_{1out}(t-\tau)=E_{2in}(t)$ with $\tau$ denoting the
propagation delay and  the corresponding parameters for the sending
and receiving nodes are same. Assuming $g=1.6\times10^{10}\,
m^{1/2}s^{-1}$, $P=100$,
$\tilde{E}_{out}(t)=i\sqrt{\frac{\sqrt{2}}{a\sqrt{\pi}}}\,e^{-(ct/a)^2}
m^{-1/2}$ with $c=1.5\times10^8$ m/s and $a=0.3$ m, in figure
\ref{fig:2}, we illustrate the transfer of a qubit from node 1 to
node 2 by the mapping transformation
$(\alpha_g|g\rangle+\alpha_s|s\rangle)|g\rangle\rightarrow|g\rangle(\alpha_g|g\rangle+0.9950\alpha_s|s\rangle)$
with $\beta_{s1}(\infty)=0.0095$. The fidelity of this operation is
$F=0.9900$ for the transferred state with coefficients
$\alpha_g=\alpha_s=1/\sqrt{2}$. If $P=1000$ with other parameters
unchanged, we have $F=0.9990$. Using the same parameters as those
used in the figure \ref{fig:2}, we present in figure \ref{fig:3} the
creation of entanglement of the qubits in neighboring nodes through
the transformation:
$|s,g\rangle\rightarrow0.7047|s,g\rangle+0.7004|g,s\rangle$.  The
target mapping is
$|s,g\rangle\rightarrow1/\sqrt{2}(|s,g\rangle+|g,s\rangle)$, so the
fidelity of this operation is $F=0.987$. Note that in the receiving
node, the control field $\Omega_2(t)$ must be designed to absorb a
whole photon, no matter whether the incoming field contains a whole
 photon or not. Otherwise, the operation of either the state transfer or the generation of entanglement
  between two nodes will yield wrong result.

\begin{table}
\caption{\label{tab1}Effect of errors in parameters on the fidelity
of state transfer operations. The parameters remain the same as
those used in figure \ref{fig:2}. When there are no errors, the
fidelity $F=0.9900$.}
\begin{ruledtabular}
\begin{tabular}{ccccc}
node1&10\% $g$ err&10\% $\Gamma_{pl}$ err&10\% $\Gamma'$ err&10\% $\Omega_1(t)$ err\\
Fidelity&0.8910&0.9429&$0.9895$&$0.9845$\\
\hline node2&10\% $g$ err&10\% $\Gamma_{pl}$
err&10\% $\Gamma'$ err&10\% $\Omega_2(t)$ err\\
Fidelity&0.8940&0.9426&$0.9895$&$0.9840$
\end{tabular}
\end{ruledtabular}
\end{table}

In the above analysis, exact knowledge of the parameters is assumed.
Table \ref{tab1} shows the effect of the unknown errors in the
various parameters on the fidelity of the transfer of a qubit from
the sending node to the receiving one. From Table \ref{tab1}, we see
that the scheme is robust against experimental imperfections except
the uncertainty in coupling $g$ and the resulting $\Gamma_{pl}=2\pi
g^2/c$,  which can be overcome, since the position of the emitter on
which the coupling $g$ is dependent \cite{dcasp} can be determined
with very high accuracy. $\Omega(t)$ can also have unknown phase
error due to laser fluctuation which can be considered static in the
time scale of ns. The problem resulting from the relative phase
between $\Omega_1(t-\tau)$ and $\Omega_2(t)$ can be solved by
employing  a delayed phase locking of the control field in the two
nodes \cite{wang}.

\section {IV. CONCLUSION}
We have shown a general and exact solution  for the dynamics of a
quantum interface consisting of a three-level optical emitter
coupled to a proximal nanotip. The scheme enables many essential
quantum network operations including sending, receiving, swapping,
and entangling qubit at distant nodes to be performed with
  near  unity fidelity and in a  shorter operation time than those
 based on the cavity QED. Furthermore, the setup is easier to be miniaturized, thus
 would significantly facilitate the scalability of quantum computers.
 This scheme is applicable to a wide range of physical
 implementations of quantum interface such as solid state systems \cite{beka},
 trapped ions \cite{phys1} and  ultrasmall quantum dots\cite{pra57,phys4}
with discrete levels and in particular with electron spin levels.
\section {ACKNOWLEDGMENTS}
This work was supported by the State Key Programs for Basic Research
of China (2005CB623605 and 2006CB921803), and by National Foundation
of Natural Science in China Grant Nos. 10474033 and 60676056.


\begin{references}
\bibitem{jcae}J.I. Cirac, A.K. Ekert, S.F. Huelga, and C. Macchiavello,  Phys. Rev. A {\bf 59}, 4249 (1999).
\bibitem{ddiv}D. DiVincenzo, Fortschr. Phys. {\bf 48}, 771 (2000).

\bibitem{rtgr} R.J. Thompson, G. Rempe, and H.J. Kimble, Phys. Rev. Lett. {\bf68}, 1132 (1992).
\bibitem{mbru} M. Brune {\it et al.}, Phys. Rev. Lett. {\bf76}, 1800 (1996).
\bibitem{cirac} J.I. Cirac, P. Zoller, H.J. Kimble, and H. Mabuchi, Phys.
Rev. Lett. {\bf78}, 3221 (1997).
\bibitem{adia} L.-M. Duan, A. Kuzmich, and H. J. Kimble, Phys. Rev. A {\bf 67},
032305 (2003).
\bibitem{wang} W. Yao, R.-B. Liu, and L.J. Sham, Phys. Rev. Lett. {\bf95},
030504 (2005).
\bibitem{lcas} L. Childress, A.S. S{\o}rensen, and M.D. Lukin, Phys. Rev. A {\bf69}, 042302 (2004).
\bibitem{assr} A.S. S{\o}rensen {\it et al.}, Phys. Rev. Lett. {\bf92}, 063601 (2004).
\bibitem{abla} A. Blais {\it et al.}, Phys. Rev. A {\bf69}, 062320 (2004).
\bibitem{awal} A. Wallraff {\it et al.}, Nature (London) {\bf431}, 162 (2004).
\bibitem{dcas} D.E. Chang, A.S.S{\o}rensen, P.R. Hemmer, and M.D. Lukin,  Phys. Rev. Lett. {\bf97}, 053002 (2006).
\bibitem{dcass} D.E. Chang, A.S.S{\o}rensen, E.A. Demler, and M.D. Lukin, Nat. Phys. {\bf3}, 807 (2007).
\bibitem{mist}M.I. Stockman, Phys. Rev. Lett. {\bf93}, 137404 (2004).
\bibitem{crcn}C. Ropers, C.C. Neacsu, T. Elsaesser, M. Albrecht, M.B. Raschke, and C. Lienau, Nano Lett. {\bf7}, 2784 (2007).
\bibitem{dcasp} D.E. Chang, A.S.S{\o}rensen, P.R. Hemmer, and M.D. Lukin, Phys. Rev. B {\bf76}, 035420 (2007).
\bibitem{gsjk} G. Schider, J.R. Krenn, A. Hohenau, H. Ditlbacher, A. Leitner, F.R. Aussenegg, W.L. Schaich, I. Puscasu, B. Monacelli, and G. Boreman, Phys. Rev. B {\bf68}, 155427 (2003).
\bibitem{dsss} D.B. Strasfeld, S.-H. Shim, and M.T. Zanni, Phys. Rev. Lett. {\bf99}, 038102 (2007).
\bibitem{inag} I. Novikova, A.V. Gorshkov, D.F. Phillipps, A.S. S{\o}rensen, M.D. Lukin, and R.L. Walsworth, Phys. Rev. Lett. {\bf98}, 243602 (2007).
\bibitem{mpsm} P. Meystre and M. Sargent III, Elements of Quantum Optics 3rd edn (Springer, New York, 1999).
\bibitem{lryw} R.B. Liu, W. Yao, and L.J. Sham, Phys. Rev. B {\bf72}, 081306 (2005).
\bibitem{mtag} M.J. Testolin, A.D. Greentree, C.J. Wellard, and L.C.L. Hollenberg, Phys. Rev. B {\bf72}, 195325 (2005).
\bibitem{rhle} R.H. Lehmberg, Phys. Rev. A {\bf 2}, 889 (1970).
\bibitem{beka}B.E. Kane, Nature {\bf 393}, 133 (1998).
\bibitem{phys1} J.F.Poyatos, J.I.Cirac and P.Zoller, Fortschr.Phys {\bf 48},
785 (2000).
\bibitem{pra57} D.Loss and D.P.DiVincenzo, Phys. Rev. A {\bf 57},
120 (1998).
\bibitem{phys4} G.Burkard, H.-A.Engel and D.Loss, Fortschr.Phys {\bf 48}, 965
(2000)

\bibitem{kpna} K.P. Nayak, P.N. Melentiev, M. Morinaga, F.L. Kien, V.I. Balykin, and K. Hakuta, Opt. Express {\bf15}, 5431 (2007).
\end{references}
\end{document}